\documentclass[aps,superscriptaddress,twocolumn]{revtex4-1}
\usepackage{graphicx}
\usepackage{dcolumn}
\usepackage{bm}
\usepackage{colordvi}
\usepackage{color}

\begin{document}

\title{Spin rotation by resonant electric field in few-level quantum dots:  Floquet dynamics and tunneling}
\author{D.V. Khomitsky}
\email{khomitsky@phys.unn.ru}
\affiliation{Department of Physics, National Research Lobachevsky State University of
Nizhni Novgorod, 603950 Gagarin Avenue 23, Nizhny Novgorod, Russian
Federation}
\author{E.A. Lavrukhina}
\affiliation{Department of Physics, National Research Lobachevsky State University of
Nizhni Novgorod, 603950 Gagarin Avenue 23, Nizhny Novgorod, Russian
Federation}
\author{E.Ya. Sherman}
\affiliation{Department of Physical Chemistry, The University of the Basque Country,
48080 Bilbao, Spain}
\affiliation{IKERBASQUE Basque Foundation for Science, Bilbao, Spain}

\begin{abstract}

We study electric dipole spin resonance caused by sub-terahertz (THz) radiation in a multilevel finite-size quantum dot formed in a nanowire
focusing on the range of driving electric fields amplitudes where a strong interplay between the Rabi spin oscillations and tunneling
from the dot to continuum states can occur. 
A strong effect of the tunneling on the spin evolution in this regime occurs due to formation of mixed spin states.
As a result, the tunneling strongly limits possible spin manipulations time.  
We demonstrate a backaction of the spin dynamics on the tunneling and position
of the electron. The analysis of the efficiency of the spin manipulation in terms of the system energy shows 
that tunneling decreases this efficiency. Fourier spectra of the time-dependent expectation value of the electron position 
show a strong effect of the spin-orbit coupling on their low-frequency components. This results can be applied to operational
properties of spin-based nanodevices and extending the range of possible spin resonance frequencies to the THz domain.

\end{abstract}

\date{\today}
\maketitle

\section{Introduction}

Implementation of electron spins in nanoscale semiconductors such as quantum dots and other 
heterostructures as scalable hardware elements for quantum 
information processing \cite{Loss1998,Burkard1999} requires tools for fast and efficient spin manipulation
in these systems. A preferable way of such manipulations is related to application by electric rather 
than by  magnetic fields since well-controllable electric field can be produced on demand by modern
nanoscale electrical engineering tools. The venue for such a manipulation is offered either by the natural or by 
engineered presence of spin-orbit coupling (SOC) in these semiconductor structures, leading to the 
electric dipole spin resonance (EDSR) \cite{Rashba1960,Bemski1960,Bell1962,Bratashevskii1963}, 
and making it an efficient tool of manipulation of electron spins in solids. This ability 
to manipulate spin by electric field appears, in general, due to the fact that spin precession 
is related to the electron displacement caused by application of external 
electric field \cite{Ramsak2014}. 

There are three basic systems where the electric dipole spin resonance can be observed \cite{Rashba2018}. 
First realization is given by itinerant electrons in bulk crystals \cite{Rashba1960,Bemski1960,Bell1962,Bratashevskii1963}, 
surfaces \cite{Azpiroz} with spin-orbit coupling \cite{Vasko1980},  and in artificial semiconductor structures such as heterojunctions and quantum wells \cite{Rashba2003}
and superlattices \cite{Khomitsky2008}. 
Second realization, theoretically analyzed in 
Refs. \cite{Golovach2006,Jiang2006,Hu2012,KGS2012,Borhani2012,Nowak2012,Ban2012,Li2013,Romhanyi2015,
Veszeli2018,Stipsic2020} and experimentally realized for electrons \cite{Nowack2007}  
and holes \cite{Studenikin2019},  is related to carriers in single and double semiconductor quantum dots 
(a similar approach can be applied for carriers in carbon nanotubes \cite{Osika2015}). 
Due to a strong confinement by external potential, here the 
spectrum of carriers is discrete, and the carrier always remains in the dot, either two-
\cite{Nowack2007} or one-dimensional \cite{Nadj2010}, even
when a strong electric field is applied. Third realization occurs for 
potentials vanishing at large distances with the examples being: 
electrons (holes) on donors (acceptors) \cite{Rashba1964a,Rashba1964b,Linpeng2016,Linpeng2018}, 
in finite-range potential produced by various charged gates \cite{Davies1995,Malshukov2003,Serra2005,Zibold2007,Sadreev2015}, 
and in surface-based self-assembled quantum dots \cite{Sheng2006,Takahashi2010,Gawarecki2018}. 

An important example of spin manipulation in system with the confining potential vanishing at 
large distances is given by gate-defined quantum dots- \cite{Caroll2019} and P donors in Si \cite{Caroll2015},
where the hyperfine coupling to donor nuclear spins results in experimentally achieved possibilities of spin qubit manipulations 
\cite{Morello2018,Weber2018,Morello2018b,Lyon2019,Morello2020}. In a spin ensemble of these donors, the coherent Rabi dynamics 
has been observed  [\onlinecite{Lyon2017}]. However, the effect of SOC in these systems is weaker than that achievable
in other nanostructures such as InSb quantum wires and dots and other materials with large tunable SOC. 
Since the large SOC generally leads to faster spin flip and shorter operating times, 
these semiconductor structures are of interest for the applications, and will be considered here in detail.

This third realization is of our interest here since low-frequency electric
fields can ionize the electron states by tunneling and, thus, cause additional nontrivial 
spin dynamics resulting from the spatial spread of the electron wavefunction.
For example, even for the inter-minima tunneling in a double quantum dot at relatively short 
distances the coupled spin and spatial dynamics may cause strong nonlinearities in the spin precession 
Rabi frequency dependence on the driving field amplitude \cite{KGS2012}. 

Thus, we are interested in Floquet dynamics in a tunneling system, where a combination of two long 
time scales becomes important. The first long time scale is the time 
of spin flip driven by electric field, inversely proportional to the field amplitude.
The second one is the tunneling time due to driving by electric field producing coupling 
of localized stats to the continuum, being strongly, approximately exponentially,
dependent on the inverse field amplitude. If both these times are of the same order of magnitude,
the  coupled spin and charge density dynamics, studied in this paper, becomes highly nontrivial.
In a multilevel quantum dot the tunneling is a well-defined semiclassical process, in contrast to 
that in shallow quantum dots \cite{KLS2019}. As a result of a stronger confinement, 
the amplitude of the electric field required to operate a spin flip is considerably 
larger and, therefore, the involvement of the continuum states is qualitatively different. 

\begin{figure}[tbp]
\centering
\includegraphics*[width=0.45\textwidth]{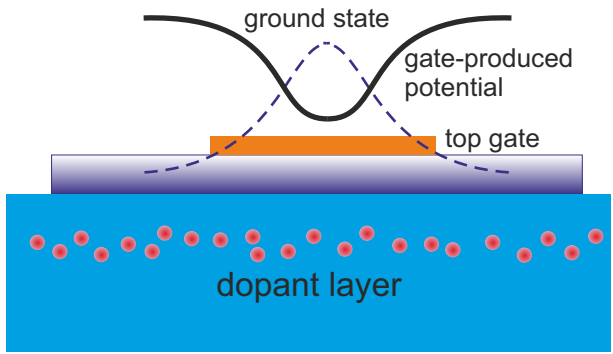}
\caption{Schematic presentation of a structure including a gated nanowire on a doped bulk substrate.}
\label{figstructure}
\end{figure}

A relevant issue here is the range of the frequencies 
required for the spin resonance. The major part of experimental setups produces
frequencies within the tens GHz-range. However, 
fast and decoherence-robust spin manipulation may require higher,  
sub-THz frequencies. The reason for this request is due to the fact that the  
possible spin Rabi frequencies decrease with the decrease in the Zeeman splitting, and,
therefore will require stronger driving fields. 
Combination of progress in development of THz sources  \cite{GanichevPrettl,Takeda2018,Singh2020} 
and demand for high driving frequencies
for the EDSR in two-dimensional systems \cite{Brooks2020} and for fast spin manipulation in conventional nanostructures will make these techniques applicable in spin-manipulation technologies.

An interesting aspect of this dynamics is that usually the EDSR assumes a 
pure spin state, defined on the Bloch sphere, where the spin rotation occurs with
a minor change in the electron orbital state. 
However, in the presence of spin-orbit coupling the spin states become mixed.  This circumstance
strongly modifies the entire spin evolution. For applications, 
the efficiency of the spin-flip process, that is the ratio 
of the Zeeman energy to the total energy acquired by electron as a result of the electric 
field action, is of interest. We will demonstrate how the efficiency depends on the system parameters. 
Moreover, the spin-orbit coupling causes a feedback of the spin dynamics 
on the tunneling efficiency and resulting electron position. Even at a relatively weak
spin-orbit coupling this effect can be strong, as we will demonstrate below. 

This paper is organized as follows. In Sec. II we introduce the Hamiltonian of our system and discuss 
time-independent and periodic contributions to the electron energy. 
In Sec. III  we describe the computational model of electron 
states and their dynamics and provide the specific system parameters. In Sec. IV the main 
results are presented: spin dynamics, structure of the electron wavefunction, 
time dependencies of mean energy and efficiency of spin flip, localization probabilities, 
coordinate evolution, and its Fourier power spectra. Finally, in Sec. V we give our conclusions
and make relation to the design of  devices using the effects described in the paper.

\section{Hamiltonian and observables}

\begin{figure}[tbp]
\centering
\includegraphics*[width=0.42\textwidth]{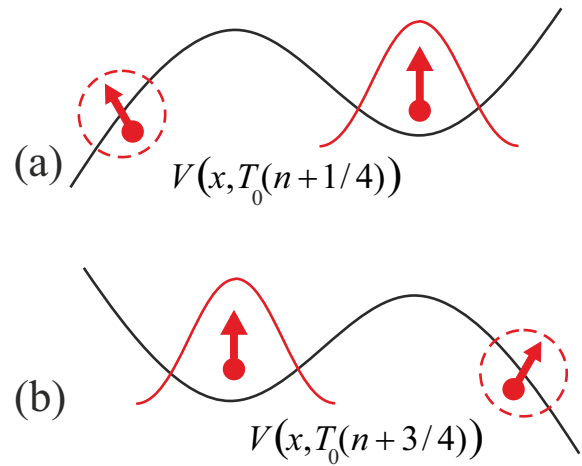}
\caption{Schematic illustration of combined tunneling and spin precession caused by a periodic 
external field. Arrows in the circles located at the points of tunneling-induced escape 
show that dependent on the direction of the tunneling,
spin precesses in opposite directions.}
\label{figmqw}
\end{figure}

We consider a narrow nanowire, elongated along the $x-$ axis located on the top of a doped substrate,
as schematically presented in Fig. \ref{figstructure}. Assuming that the transversal modes are not excited,
we characterize the electron motion by a 
two-component wave function ${\bm\psi}(x,t)=[\psi_{1}(x,t),\psi_{2}(x,t)]^{\rm T},$  where ${\rm T}$
stands for the transposition, and use the effective mass approximation
Hamiltonian with: 
\begin{equation}
H(t)=\frac{\hbar^{2}k^{2}}{2m}+V(x,t)+\frac{\Delta}{2}\sigma_{z}+\alpha \sigma_{y}k. 
\label{h1}
\end{equation}
Here $m$ is the electron effective mass, 
and $k=-i\partial/\partial x$ is the wave vector operator. 
The time-dependent potential is the sum of the static confinement potential $U(x)$, as can be produced 
by electrostatic gate shown schematically by a rectangle above the nanowire in Fig. \ref{figstructure},
and  external driving giving 
\begin{equation}
V(x,t)=U(x)+Fx\sin \omega_{d} t,
\label{Vxt}
\end{equation}
where $F\equiv e{\cal E}_{0}$, $e$ is the fundamental charge and ${\cal E}_{0}$ is the field amplitude.  
The Rashba SOC $H_{R}=\alpha \sigma _{y}k$ while, more complicated forms of spin-orbit coupling can be found
in Refs. [\onlinecite{Mireles2001,Gelabert2010,Zainuddin2011,Xu2014}]. The constant
magnetic field $\mathbf{B}=(0,0,B)$ produces the Zeeman term in the
Hamiltonian. Here the Zeeman splitting $|\Delta|=\mu_{B}|gB|,$ where $\mu_{B}$ is the Bohr magneton, 
and $g$ is the Land\'{e}-factor. The spin-split Zeeman partners 
participate in the spin resonance driven by the external periodic electric field in (\ref{Vxt}).
The reason for this participation is that  due to the presence of SOC
the eigenstates of (\ref{h1}) at $F=0$ always contain both spin components.
Therefore, spin-flip transitions still be caused by the electric field
described with the position operator, producing the EDSR. 
In this case, the well-defined spin Rabi oscillations occur if one neglects the tunneling
effects, that is coupling to the other orbital and continuum states. 
The dynamical interaction between discrete and continuum states during the evolution 
driven by periodic electric field causes nontrivial effects to be discussed 
in the following Sections. 
One of the effects is illustrated schematically in Fig. \ref{figmqw}: in the presence of 
SOC the direction of spin precession is different, depending on the direction of 
the tunneling escape, making a controlled spin flip rather challenging task especially 
if the tunneling is triggered by the alternating field responsible for the spin evolution.

To describe the evolution we solve the nonstationary Schr\"{o}dinger equation
\begin{equation}
i\hbar\partial_{t}{\bm\psi}(x,t)=H(t){\bm\psi}(x,t), 
\label{scheq}
\end{equation}
where $\partial_{t}\equiv \partial/\partial t,$ and calculate expectation values 
for experimentally measurable observables ${\cal O}$ as:
\begin{equation}
\langle{\cal O}(t)\rangle=\int_{-\infty}^{\infty}{\bm\psi}^{\dagger}(x,t){\cal O}{\bm\psi}(x,t) dx. 
\label{observable}
\end{equation}
The observables will be specified later in the text. 

\section{Model and numerical approach}

\subsection{Basis states and model of the dynamics}

We choose the static potential $U(x)=-U_{0}/\cosh^{2} (x/d),$ assuming that it is deep such that 
the parameter $\xi\equiv U_{0}md^{2}/\hbar^{2}\gg1.$ Although  the exact bound states for this potential
are well-known analytically \cite{Landau} in terms of the Legendre functions, 
it is practical to consider simple approximations based on the expansion at $x\ll\,d$ as
$U(x)=-U_{0}\left(1-x^{2}/d^{2}\right).$
As a result, the lowest bound states can 
be described in terms of a harmonic oscillator with the energy interval $\hbar\omega_{\rm ho}=U_{0}/\sqrt{\xi/2}\ll U_{0}$
and the Gaussian width $d/(2\xi)^{1/4}\ll d.$ The numerically accurate 
eigenenergies $E_{n}^{(0)}$ and basis states of the Hamiltonian $\hbar^{2}k^{2}/2m+U(x)$ 
are found by expansion in the basis of periodic functions $\sin(x/\lambda_{n}+\delta_{n})$ where $\lambda_n= 2L/(\pi n)$ and $\delta_n=\pi n /2$, $n=1,2, \ldots$ satisfying zero boundary conditions
at $x=\pm L,$ where $2L$ is a large wire length.  

Then, we numerically produce basis of two-component eigenstates ${\bm\phi}_{n}(x)$ of the spinful Hamiltonian
\begin{equation}
H_{0}=\frac{\hbar^{2}k^{2}}{2m}+U(x)+\frac{\Delta}{2}\sigma_{z}+\alpha \sigma_{y}k 
\label{h0}
\end{equation}
with the spin-split eigenenergies $E_{n}$ such as $H_{0}{\bm\phi}_{n}(x)=E_{n}{\bm\phi}_{n}(x)$. 
As a result, the evolution of wave function is presented as
\begin{equation}
{\bm\psi}(x,t)=\sum_{n}c_{n}(t)e^{-iE_{n}t/\hbar}{\bm\phi}_{n}(x). 
\label{wf}
\end{equation}
Thus, the problem is reduced to obtaining full set of coefficients $c_{n}(t)$ by numerical solution
of a large system of linear differential equations, as will be presented below with some computations details
given in Ref. [\onlinecite{KLS2019}]. Here we choose the interval $2L=320d$ and take 4000 
basis states ${\bm\phi}_{n}(x).$ The initial state is chosen as ${\bm\psi}(x,0)={\bm\phi}_{1}(x),$ 
that is the ground state of $H_{0},$ where for the realistic parameters one has $\langle\sigma_{z}(0)\rangle\approx 1.$  

\subsection{System parameters}

In the following analysis we consider a gate-formed quantum dot based on InSb quantum wire, see Fig. \ref{figstructure}. 
For this material $m=0.0136$ of a free
electron mass \cite{Saidi} and $g=-50.6.$ The magnitude of SOC $\alpha $ in InSb can be tuned by the gate
voltage up to 100 meVnm \cite{Leontiadou,Wojcik}. 

For the quantum dot parameters we accept $U_{0}=27$ meV and $d=50$ nm with 
five discrete quantum dot levels $E_{1}^{(0)},\ldots , E_{5}^{(0)}$ being formed \cite{Landau} 
with $\omega_{0}=(E_{2}^{(0)}-E_{1}^{(0)})/\hbar = 13.43 \mbox{ ps}^{-1}$, 
and the associated period $T_{0}=2\pi/\omega_{0}=0.468$ ps gives the natural time scale for the evolution.
We consider the magnetic field $B=0.447$ T, providing the Zeeman 
splitting of the ground state corresponding to the resonant driving frequency 
(including the contribution of spin-orbit coupling) $\omega_{d}=2 \mbox{ ps}^{-1}$ (0.32 THz), 
giving the ratio to the fundamental frequency $\omega_{d}/\omega_{0}=0.149$. 
Such Zeeman splitting requires a sub-THz-range driving hardware, as it has been discussed in the Introduction. 
The advantages of this large splitting come from the following circumstances: it is known that smaller 
Zeeman splitting gives smaller values of spin-flip matrix elements of the nonstationary part in  (\ref{Vxt}) \cite{KLS2019}, 
which results in a higher driving field amplitude $F$ required to operate the spin flip in a desired time. 
As a result, the leakage to the continuum will be strongly enhanced, as we show below. 
Thus, the larger driving frequencies may provide the desired spin flip time with lower operating 
field which fosters the applicability of the proposed mechanism.

We consider the effect of a relatively weak SOC $\alpha=5$  meVnm and a relatively 
strong $\alpha =25$ meVnm for various driving amplitudes. Taking into account the material parameters
for spin-obit coupling produced by external electric field in InSb \cite{Winklerbook}, we obtain that the  
effective two-dimensional donor dopant layer concentration, as shown in Fig. \ref{figstructure}, 
corresponding to this spin-orbit coupling,  is in the range of $10^{11}-10^{12}$ cm$^{-2}.$
Since the nonuniform static electric field forming the quantum dot has a $z-$ component, it can 
contribute to the Rashba coupling. Taking into account that this field component cannot exceed $\sim U_{0}/ed,$ and using the 
corresponding material parameter \cite{Winklerbook}, we obtain that this contribution 
to the SOC coupling is less than 2.5 meVnm, considerably smaller than the accepted values for InSb nanowire. 

Now we discuss the effect of the Dresselhaus SOC, typical for III-V semiconductors.
The corresponding Hamiltonian has the form $H_{D}=\alpha_{D}\left({\bm\kappa}{\bm \sigma}\right)$, where $\alpha_{D}$
is the coupling constant and ${\kappa}_{x}=k_{x}\left(k_{y}^{2}-k_{z}^{2}\right),$ with other components obtained
by cyclic permutation. This ${\bm\kappa}-$ dependence makes the Dresselhaus SOC in nanowires strongly dependent
on their shape and orientation. For example, for the wire grown along the crystallographic $x$-axis, we obtain
$H_{D}^{[x]}=\alpha_{D}k_{x}\left(\langle k_{y}^{2}\rangle-\langle k_{z}^{2}\rangle\right)$, where $\langle k_{i}^{2}\rangle$
is the shape-dependent expectation value of the corresponding operator
\cite{Dyakonov1986,Rashba1986}. As a result, this coupling can be made 
zero for the highly symmetric wires of the appropriate cross-section. For this growth direction, the Dresselhaus coupling is similar to the 
Rashba coupling in Eq. (\ref{h0}). For the wires grown along the $y$ or $z$-axis, 
it will modify only the direction of the SOC field without considerable 
changes in the spin dynamics. 
Taking into account the estimate $\alpha_{D}\approx 0.76\mbox { eVnm}^{3}$ \cite{Winklerbook}
we obtain for typical values $\left(\langle k_{y}^{2}\rangle - \langle k_{z}^{2}\rangle\right)\sim 10^{-2}\mbox{ nm}^{-2}$ 
the coupling of the order of 10 meVnm, not considerably modifying the parameters of the driving fields of interest
for the present work.

\section{Spin dynamics and feedback on the tunneling and position}

\subsection{Time dependence of spin: the role of the tunneling}

\begin{figure}[tbp]
\centering
\includegraphics*[width=0.48\textwidth]{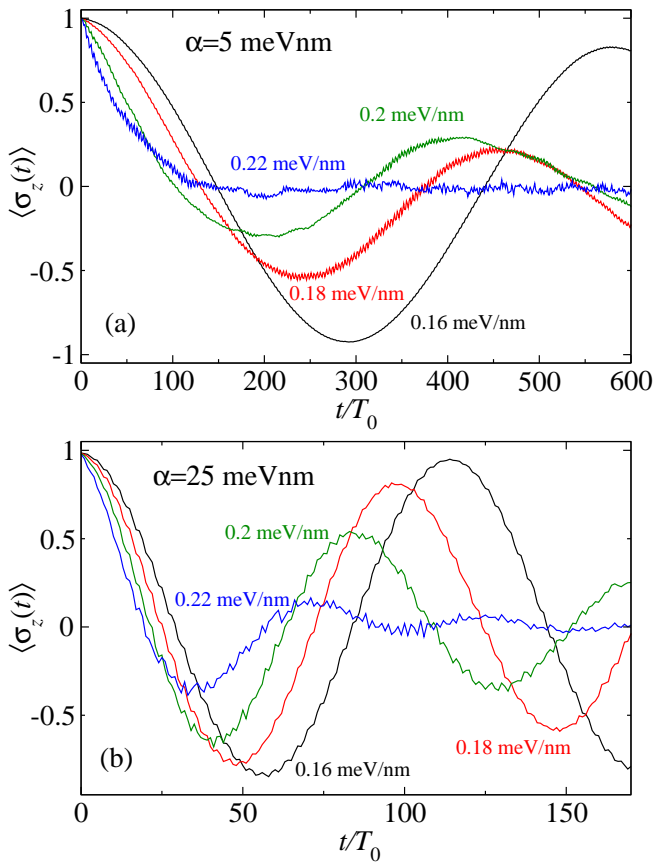}
\caption{Spin projection $\langle\sigma_{z}(t)\rangle$ for $\alpha=5$ meVnm (a) and 
$\alpha=25$ meVnm (b) for 
various driving field amplitudes as marked near the curves. At a relatively high $F \ge 0.2$ meV/nm the spin 
dynamics demonstrates visible damping of the Rabi oscillations, as caused by the interaction 
with continuum states increasing with the driving field amplitude.}
\label{figsz}
\end{figure}

In Fig. \ref{figsz} we show the time dependence of $\langle\sigma_{z}(t)\rangle$ corresponding 
to Eq. (\ref{observable}) with ${\cal O}={\sigma}_{z}$ giving
\begin{equation}
\langle \sigma_{z}(t) \rangle=\sum_{n_{1},n_{2}}c^{*}_{n_{1}}(t) c_{n_{2}}(t) e^{-i(E_{n_{2}}-E_{n_{1}})t/\hbar}
\langle{\bm\phi}_{n_{1}}|{\sigma}_{z}|{\bm\phi}_{n_{2}}\rangle. 
\label{sigmaz}
\end{equation}
Here and below the time is measured in the units of $T_{0}$, and we 
track the evolution on calculation time scales of about $650 T_{0}$ 
for small SOC and $200 T_{0}$ for large SOC. These ranges are determined by the spin flip process 
which needs to be well-developed. As a result, one can obtain from the spin dynamics the characteristic 
spin evolution time. One can see that a sizable decrease in this time can be observed for moderate increase 
of $F$ from $0.16$ to $0.2$ meV/nm which is in the framework of usual Rabi frequency dependence on 
the driving field amplitude. When the field grows further, the electron escapes the quantum 
dot quickly, before well-established Rabi spin  
oscillations are developed. This effect is stronger at lower values of SOC $\alpha=5$ meVnm. 
One can see also that at the fields $F \ge 0.2$ meV/nm the Rabi oscillations are accompanied 
by visible damping. We attribute this effect to the interactions between localized and continuum states 
which is enhanced at higher driving fields. The continuum states form a dense set of 
levels with alternating spin projections due to the Zeeman splitting \cite{KLS2019}. Transitions 
to these states driven by external electric field lead to the wavefunction consisting of 
spin-up and spin-down states, and their relative contributions to the total norm 
have a tendency to equalize when the electron is pushed into the continuum. 

To understand the interplay of tunneling and spin flip we first 
evaluate the upper limit for the semiclassical tunneling
rate $w_{\rm tun}$ in the model triangular potential $Fx$
neglecting tunneling from the excited states as { \cite{Delone}: }
\begin{equation}
\ln \frac{\omega_{0}}{w_{\rm tun}}<\frac{4\sqrt{2}}{3}\xi^{1/2}\frac{U_{0}}{Fd}. 
\label{tunneling1}
\end{equation}
Equation (\ref{tunneling1}) gives the upper estimate of the quantity since the real tunneling barrier
is lower and more narrow than the triangular one. 
Although this equation shows that the  tunneling probability rapidly increases at $Fd$ 
being a sizable fraction of $U_{0},$ the actual tunneling rate, very strongly $F-$ dependent,
can only be obtained numerically.  
Since spin-flip Rabi frequency $\Omega_{R}$ is a linear function of $F$ (being strictly linear when the two-level approximation is applicable)
with $\Omega_{R}T_{0}\ll 1$, the effect of tunneling on the spin-flip
processes occurs in a relatively narrow range of the driving fields when two rates are comparable. 

For a stronger Rashba coupling the spin dynamics becomes faster and well-defined 
Rabi oscillations have enough time to develop before the tunneling occurs,
as it can be seen in Fig. \ref{figsz}(b). However, 
when the field exceeds $0.2$ meV/nm, the electron tunnels from the dot so quickly that 
the spin flip has no time to be established, regardless of the spin-orbit 
coupling strength, and the continuum states play the dominant role.
In other words, high values of $F$ produce the tunneling rate $w_{\rm tun}$ (see Eq.(\ref{tunneling1})) satisfying 
the inequality $\Omega_{R}\lesssim w_{\rm tun},$  where the spin flip time is longer than the tunneling escape 
time. As to the driving field range desired for controlling spin flips, a small  
amplitude produces slow Rabi oscillations, which on long times may be hampered by various decoherence processes. 
So, it is one of our goals to 
estimate accurately the optimal range of electric fields where the driving is strong enough to trigger 
fast spin flips while being not sufficiently strong to cause the quantum dot ionization.

\begin{figure}[tbp]
\centering
\includegraphics*[width=0.48\textwidth]{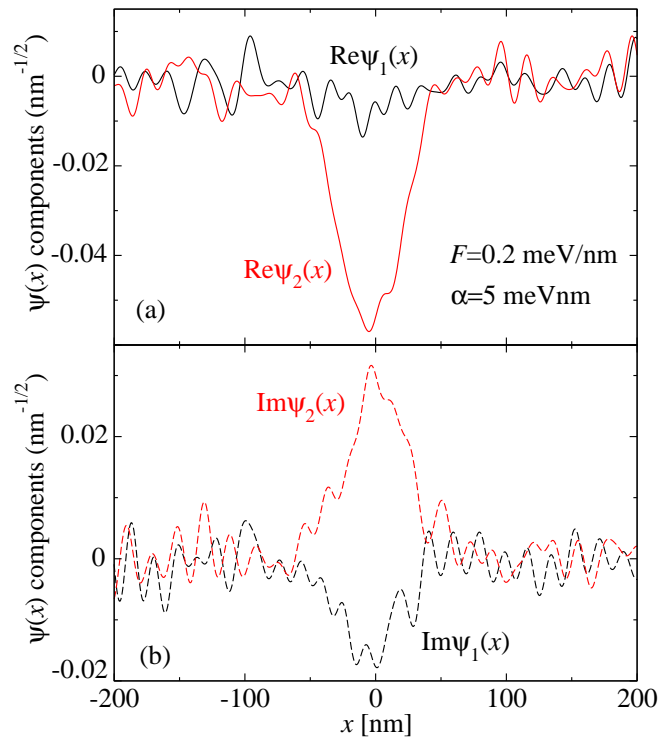}
\caption{Components of the electron spinor ${\bm\psi}(x,t)$ function at $t=655T_{0}:$ (a) real parts,
(b) imaginary parts.}
\label{mixed}
\end{figure}

It is important that the spin states we consider are mixed rather than pure. This can be 
seen with the $(2\times2)$ spin density matrix 
\begin{equation}
{\bm\rho}(t)=\int_{-L}^{L}{\bm\psi}(x,t){\bm\psi}^{\dagger}(x,t)dx. 
\label{denmat}
\end{equation}
At $t=0,$ we have $|\psi_{2}(x,0)|\ll|\psi_{1}(x,0)|$ with $\langle \sigma_{z}(0)\rangle\approx 1,$ such that
the initial spin state has a high purity with ${\rm tr}{\bm\rho}^{2}(0)\approx\,1.$ Note that 
a state is pure if and only if $\psi_{1}(x,t)=c\psi_{2}(x,t),$ where $c$ is a complex constant. 
With the course of time, two components of ${\bm\psi}(x,t)$ begin to develop 
different shapes, producing spin density matrix in Eq. (\ref{denmat}) with ${\rm tr}{\bm\rho}^{2}(t)<1$
characterizing a mixed spin state \cite{Blumbook}.  
An example of the structure of mixed spin states
in the driven spin dynamics is shown in Fig. \ref{mixed} 
at the end of the computational evolution $t=655 T_0$ for $\alpha=5$ meVnm and $F=0.2$ meV/nm. 
Although at this time and driving field, 
the electron density still has a well-defined peak inside the confining potential, a broad 
contribution of the continuum states with a relatively low total probability density 
has already been formed at $d<|x|<L,$ possessing more than half of the total norm.

This Figure shows that $\psi_{1}(x,t)$ and $\psi_{2}(x,t)$ are strongly 
different, both in the real and imaginary parts. Densities of spin components $|\psi_{1}(x,t)|^{2}$
and $|\psi_{2}(x,t)|^{2}$ (not shown in the Figure) are shifted with respect to each other,
corresponding to Fig. \ref{figmqw}, where direction of the tunneling determines the direction of the spin precession.

By evaluating the spin flip times in Fig. \ref{figsz} one may notice that they are of the 
order of $(30 \ldots 40) T_{d}$ for $\alpha=5$ meVnm and $(5 \ldots 8) T_{d}$ for $\alpha=25$ meVnm 
where $T_{d}=6.71 T_{0}$ is the driving field period. A difference to Ref. \cite{KLS2019} is that the electric field amplitude 
required to operate a spin flip in a multilevel dot is three orders of magnitude larger than that for 
the shallow dot. We attribute these differences to the considerably stronger localization and energy scales 
of a multilevel dot, where one needs larger electric fields to induce the spin flip in a desired time. 
These larger fields lead to a broader energy interval of
the continuum states contributing to the position and spin evolution.

To summarize this subsection, we mention that with the expression for spin-flip
matrix elements of coordinate \cite{Rashba1964a,Rashba1964b} for the Zeeman-split ground state 
in quantum dots of interest, one can estimate the  
tunneling-limited maximal Rabi spin frequency $\Omega_{R}^{[\max]}$ as $\Omega_{R}^{[\max]}\sim \omega_{d}d/L_{\rm so},$
where $L_{\rm so}=\hbar^{2}/m\alpha$ is the spin precession length. In general, the length $L_{\rm so}$ is determined by the joint effect of the Rashba and Dresselhaus couplings, and the existence of this maximal Rabi 
frequency is model-independent. Taking into account that for our  set of parameters at $\alpha=5$ meVnm, $L_{\rm so}\approx 10^{-4}\mbox{ cm},$ we obtain a good agreement with the numerical results 
presented in Fig. \ref{figsz}. 
Note that such a limit has to exist for all finite-potential systems, including P donors in Si. However, every 
realization needs a special analysis, which can be a topic of a further research.

\subsection{Efficiency of spin flip}

The studies of coupled spin and coordinate dynamics can be facilitated by tracking the time 
dependence of the energy-related variables such as $\langle E (t)\rangle$
defined as
\begin{equation}
\langle E (t)\rangle=\sum_{n}|c_{n}(t)|^2 E_{n} 
\label{meanen}
\end{equation}
where the sum is taken over all basis states with coefficients $c_{n}(t)$ 
in the wavefunction (\ref{wf}) for $E_{n}$ defined with (\ref{observable}) 
for the unperturbed Hamiltonian (\ref{h0}).

With the external driving, the energy is pumped into the system, and the period-averaged 
expectation value of $\langle  E(t)\rangle$ grows with time. When  $\langle E (t)\rangle$ passes the 
threshold $\langle  E(t)\rangle=0$ between the localized and continuum states, the electron 
has effectively tunneled from the quantum dot into the continuum. 

\begin{figure}[tbp]
\centering
\includegraphics*[width=0.48\textwidth]{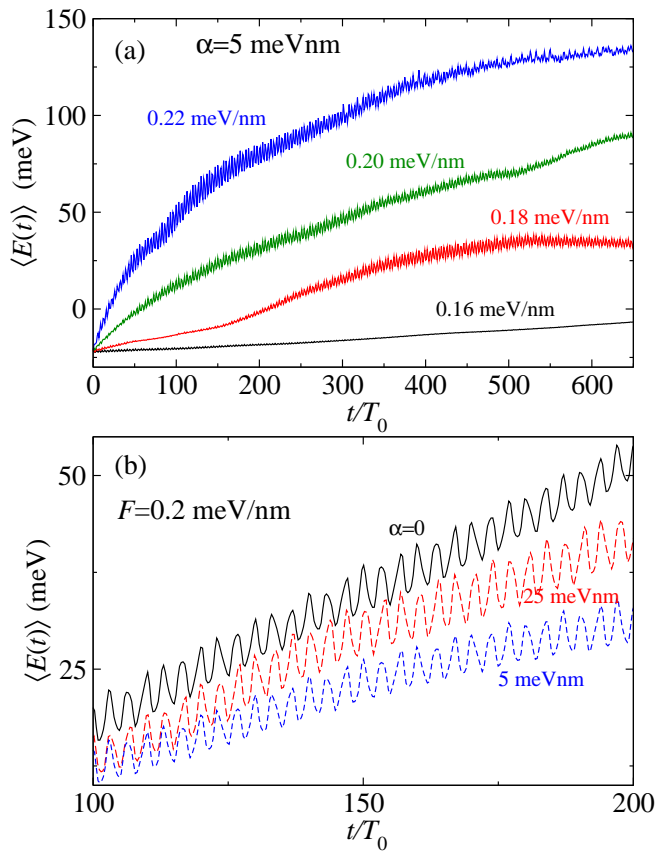}
\caption{Time dependence of mean energy $\langle E (t)\rangle$ (a) for SOC parameter  $\alpha=5$ meVnm 
and the same electric field amplitudes as in Fig. \ref{figsz}, and (b) for fixed $F_{0}=0.2$  meV/nm and different SOC 
parameters $\alpha=0$, $\alpha=5$, and $\alpha=25$ meVnm.}
\label{figen}
\end{figure}

In Fig. \ref{figen}(a) we plot $\langle E(t)\rangle$ for $\alpha=5$ meVnm 
and the same electric fields as in Fig. \ref{figsz}, and in Fig. \ref{figen}(b) 
we show examples for fixed $F=0.2$  meV/nm and different SOC parameters. 
As expected, the tunneling time quickly decreases with increasing field amplitude, and for strong fields 
greater than $0.2$ meV/nm the electron escapes into continuum during $\sim 10$ periods of driving.
The influence of SOC on the energy evolution can be illustrated by Fig. \ref{figen}(b): at the given time interval
and at zero SOC the energy grows faster than for finite SOC presented in the Figure. This effect can be attributed to 
stronger coupling of all spin-resolved $\alpha-$dependent states via the SOC, which slows to some extent the 
motion to higher energy states. 

\begin{figure}[tbp]
\centering
\includegraphics*[width=0.48\textwidth]{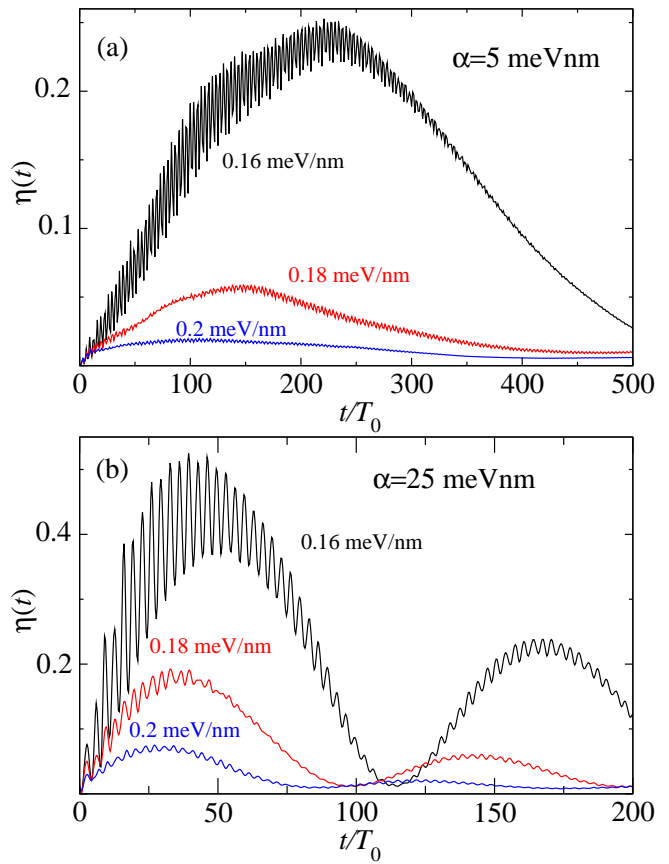}
\caption{Time dependent spin-flip efficiency (\ref{kpd}): (a) $\alpha=5$ meVnm, (b) 
$\alpha=25$ meVnm, for some of the driving fields (marking the plots) from Fig. \ref{figsz}.}
\label{figkpd}
\end{figure}

In addition to the time dependence $\langle E(t)\rangle$, it is of interest to study the spin-flip efficiency $\eta (t)$ 
defined as the ratio of the energy pumped into the spin degree of freedom to the total change of the mean energy:
\begin{equation}
\eta(t)=
\frac{\Delta}{2}\frac{\langle \sigma_{z} (t)\rangle-\langle \sigma_{z}(0)\rangle}
{\langle E (t) \rangle - \langle E(0)\rangle},
\label{kpd}
\end{equation}
where $ \langle E(0)\rangle\equiv E_{1}.$ In Fig. \ref{figkpd} we show this $\eta(t)$
for some of the electric fields from Fig. \ref{figsz}.
One can see that its time dependence is qualitatively independent of the 
SOC strength. It reaches the maximum when the first full or almost 
full spin flip is achieved. After this, the magnitude of $\eta(t)$ decreases with time. 
The explanation is straightforward: the spin dynamics and in particular the Zeeman energy have an oscillating character 
(at low driving strength) or oscillating plus decaying character (at high driving strength). So, the numerator 
in (\ref{kpd}) does not grow with time, while the denominator in general increases, as it can be seen from the
energy dependence in Fig. \ref{figen}. As to the magnitude of (\ref{kpd}), it is greater for bigger SOC 
strength since the stronger SOC coupling provides a quicker spin flip with lower corresponding change of the 
mean energy, giving the smaller denominator in (\ref{kpd}) at the moment when the maximum numerator is reached. 

\subsection{Feedback on the tunneling}

To visualize the electron escape from the quantum dot, we use the stay probability defining 
the weight of the localized eigenfunctions in the discrete part of the spectrum as:
\begin{equation}
P(t)=\sum_{i (\rm loc) } |c_{i}(t)|^{2},
\label{probloc}
\end{equation}
where summation is taken over the contribution of localized ${\bm\phi}_{i}(x)-$states. 

\begin{figure}[tbp]
\centering
\includegraphics*[width=0.48\textwidth]{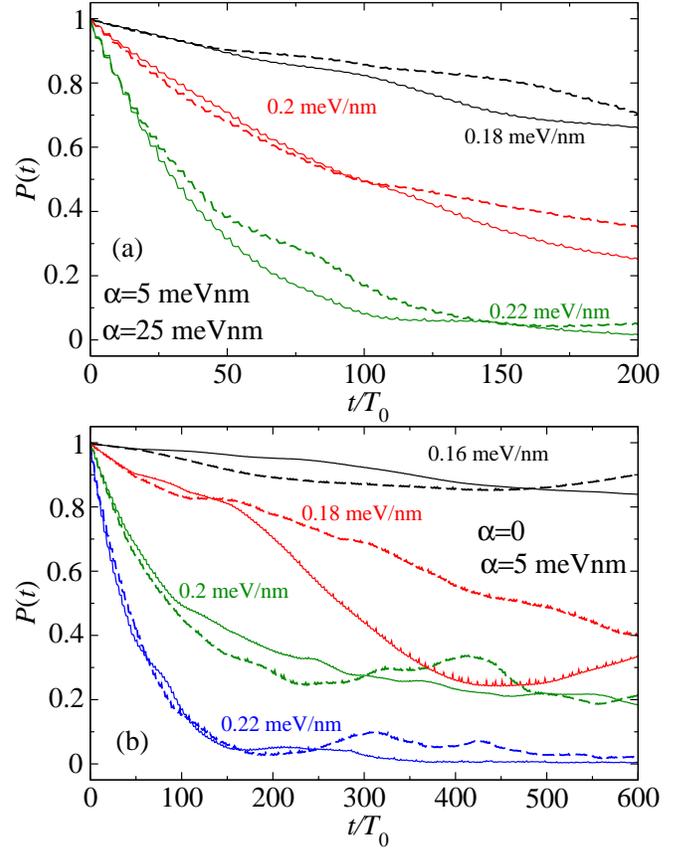}
\caption{Time dependence of localization probability (\ref{probloc}) for (a) short 
evolution times for $\alpha=5$ (dashed lines) and $\alpha=25$ (solid lines) meVnm 
and (b) long evolution times for $\alpha=0$ (solid lines) and $\alpha=5$ (dashed lines) meVnm. 
The plots are marked with the values of the driving field amplitudes.}
\label{figprobloc}
\end{figure}

The localization probability (\ref{probloc}) is shown in Fig. \ref{figprobloc} 
for two time intervals. One can see that for short evolution times shown in Fig. \ref{figprobloc}(a) both values of SOC strength 
produce similar evolution $P(t),$ and the values of (\ref{probloc}) 
for smaller $\alpha$ are in general slightly higher than for the bigger one. 
The situation is, however, somewhat different on long times to $600 T_{0}$ shown in Fig. \ref{figprobloc}(b) 
for small value of $\alpha=5$ meVnm together with the $\alpha=0$ realization. Here two $P(t)-$curves for a given 
driving strength and different SOC swap more frequently than in Fig. \ref{figprobloc}(a).

An important element for understanding of the feedback effect of spin motion on the time-dependent 
position and tunneling is the anomalous \cite{Adams} spin-dependent
velocity \cite{Landau}:
\begin{equation}
v_{\rm so}=\frac{i}{\hbar}\left[\alpha \sigma _{y}k,x\right]=\frac{\alpha}{\hbar}\sigma _{y}, 
\end{equation}
and the corresponding acceleration:
\begin{equation}
\frac{d}{dt}v_{\rm so}=\frac{i}{\hbar}\frac{\Delta}{2}\frac{\alpha}{\hbar}\left[\sigma_{z},\sigma_{y}\right]
=\frac{\alpha\Delta}{\hbar^{2}}\sigma_{x}. 
\end{equation}
This term leads to variation in the velocity due to the spin precession, as was observed in experiments 
on high-frequency conductivity of two-dimensional electron gas \cite{Tarasenko}, and the effect of time-dependent
$\langle\sigma_{y}(t)\rangle$ on $\langle x(t)\rangle$ appears as a result. The corresponding local probability 
flux satisfying equation $\partial_{t}({\bm\psi}^{\dagger}(x,t){\bm\psi}(x,t))+\partial_{x}j(x,t)=0,$
where $\partial_{x}\equiv\partial/\partial\,x,$ is given by:
\begin{eqnarray}
j(x,t)&=&
\frac{i\hbar}{2m}
\left[{\bm\psi}(x,t)\partial_{x}{\bm\psi}^{\dagger}(x,t)-
{\bm\psi}^{\dagger}(x,t)\partial_{x}{\bm\psi}(x,t)\right] 
\nonumber \\
&+&\frac{\alpha}{\hbar}{\bm\psi}^{\dagger}(x,t)\sigma_{y}{\bm\psi}(x,t). 
\end{eqnarray}
Note that for a relatively small 
coupling constant $\alpha=5 \mbox{ meVnm},$ the maximum value of the $v_{\rm so}$
velocity ${\alpha}/{\hbar}\approx 8\mbox{ nm/ps}$ can sufficiently modify the tunneling process.

\subsection{Evolution of the position}

\subsubsection{Time-dependence of the expectation values}

Along with the spin evolution, the expectation value of the coordinate $\langle x(t)\rangle$, 
is of interest for understanding of the dynamics. 
It provides information of the electron localization domain, and can help in studies 
of the tunneling effect. For example, the sufficient criteria for tunneling into the
continuum may be formulated in terms of the $\langle x (t)\rangle$ amplitude: if it 
steadily exceeds the quantum dot size $d$, the electron is out of the dot. Also, this expectation value produces 
a time-dependent electric field outside of the quantum dot, which under certain conditions, can be experimentally tracked. 

\begin{figure}[tbp]
\centering
\includegraphics*[width=0.48\textwidth]{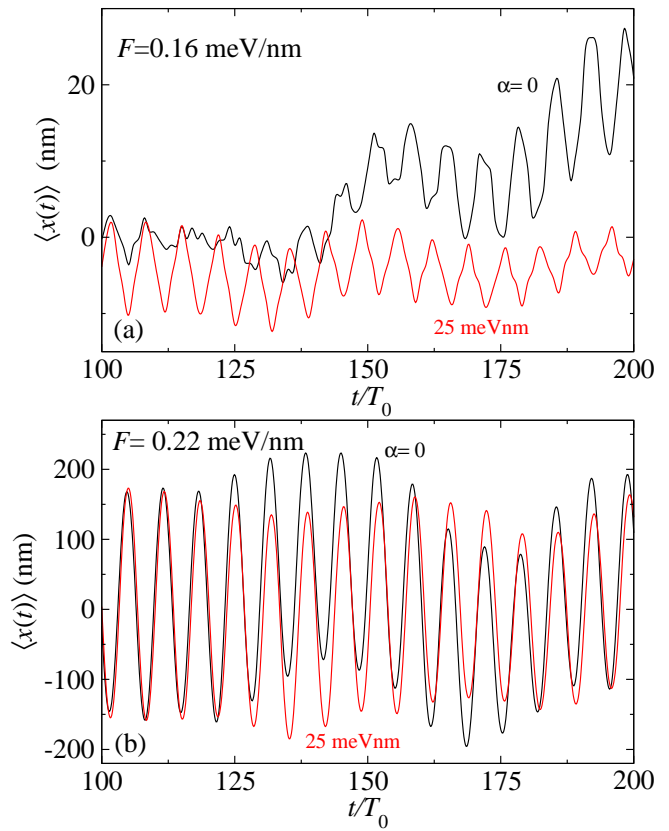}
\caption{Time dependence of mean value for the coordinate $\langle x (t)\rangle$  
for $\alpha=0$ and for high $\alpha=25$  meVnm, shown for the 
driving field amplitude (a) $F=0.16$ meV/nm and (b) $F=0.22$ meV/nm. For low 
driving amplitude $F=0.16$ meV/nm $|\langle x (t)\rangle|\ll d,$ while for $F=0.22$ 
meV/nm the amplitude of $\langle x (t)\rangle$ exceeds the quantum dot size. Here $\max\{|\langle x(t)\rangle|\}<3d,$ 
meaning that the electron motion is still strongly influenced by the $U(x)-$potential and approximately limited by the 
tunneling escape points at $\pm E_{1}/F$ where $E_1$ is the ground state energy.}
\label{figx}
\end{figure}

We calculate the dynamics of $\langle x (t)\rangle$ in analogy with (\ref{sigmaz}) where the 
spin operator is replaced by ${ x}$ with some examples of coordinate dynamics presented in Fig. \ref{figx}.  
The upper panel shows the dynamics for the low amplitude $F=0.16$ meV/nm and the bottom panel shows 
the dynamics for $F=0.22$ meV/nm. By looking in Fig. \ref{figx}(a) one can conclude that 
for low driving amplitude $F=0.16$ meV/nm the electron is still confined within the quantum dot, and its 
spin exhibits well-defined Rabi oscillations (see Fig. \ref{figsz}). When the driving amplitude grows, 
the delocalization trend is manifested, and for the highest amplitude $F=0.22$ meV/nm the 
amplitude of $\langle x (t)\rangle$ well exceeds the quantum dot size, as one can see in 
Fig. \ref{figx}(b).

One may compare the results presented in Fig. \ref{figx} with the evolution of the localization probability 
(\ref{probloc}) in Fig. \ref{figprobloc}. 
It is clear that for the given time $t$ the localization 
probability is significantly lower at higher driving fields as manifested both in small $P(t)$ 
and in large $\langle x (t)\rangle$ amplitude. Thus, the delocalization trend is tracked simultaneously 
in the coordinate and Hilbert spaces. The strong dependence of the delocalization (ionization in atomic physics) probability 
on the low-frequency electric field amplitude has an exponential character \cite{Delone} 
albeit with a rather undetermined model-dependent prefactor. Thus, our numerical findings  
shed light onto the amplitudes above which an efficient tunneling to continuum takes place. 
In addition, we note that despite two possible directions of escape in a periodic field, the probability 
density becomes broad, but remains single-peaked, as it is seen in Fig.\ref{mixed}, making  $\langle x (t)\rangle$ a reliable characteristic
of the expectation value of the electron position.

The displacements presented in Fig. \ref{figx} can be compared with the results for oscillations
amplitude $x_{0}^{\rm [loc]}$ for a particle localized in the potential $U_{0}x^{2}/d^{2}$, 
and a free electron driven by a periodic
field, $x_{0}^{\rm [cl]},$ where
\begin{equation}
x_{0}^{\rm [loc]}=\frac{d^{2}}{2U_{0}}F, \qquad x_{0}^{\rm [cl]}=\frac{1}{m\omega_{d}^2}F. 
\label{xcl}
\end{equation}
For $F=0.16\mbox{ meV/nm},$ estimated amplitude $x_{0}^{\rm [loc]}\approx 8\mbox{ nm},$ in a good agreement with
the calculations at $t<100 T_{0}$ (not shown in the Figure). Here the velocity amplitude 
$v_{0}^{\rm [loc]}=\omega_{d}x_{0}^{\rm [loc]}\approx 16\mbox{ nm/ps},$ meaning that the anomalous velocity 
$\alpha/\hbar$ plays essential role in the electron dynamics. For a delocalized electron, Eq. (\ref{xcl}) yields
for $F=0.22\mbox{ meV/nm},$ the amplitude $x_{0}^{\rm [cl]}\approx 720\mbox{ nm},$ considerably larger than the 
calculated values. This difference is due to the fact that in the tunneling ionization a broadly distributed
probability density rather than a well-defined wavepacket is being formed, resulting in a relatively small
amplitude of $\langle x(t)\rangle.$ 
This can be illustrated by the example for the wavefunction component profile 
shown in Fig. \ref{mixed} which can be compared with the localization probability in Fig. \ref{figprobloc}(b): for $F=0.2$ meV/nm
the low value of localization probability for the final moment of computation ($P\approx 0.2$) corresponds to Fig. \ref{mixed} which  
describes a wavefunction with substantial impact of continuum states demonstrating a 
widely delocalized profile rather than a well-localized wavepacket.
As to the velocity, taking into account that the numerically obtained amplitude, 
$x_{0}^{\rm [tun]}\approx 150\mbox{ nm}$ 
we find that the corresponding velocity amplitude $v_{0}^{\rm [tun]}\approx 300\mbox{ nm/ps},$ being comparable to 
the maximum anomalous velocity at $\alpha=25\mbox{ meVnm},$ confirming the importance of spin-orbit 
coupling for the electron displacement. At this point, we can define the corresponding wavevector,
$k_{0}^{\rm [tun]}\equiv v_{0}^{\rm [tun]}m/\hbar$ and the spin precession rate  
$\Omega_{0}^{\rm [tun]}=2\alpha k_{0}^{\rm [tun]}/\hbar.$ For the given set of system parameters 
and $\alpha=5\mbox{ meVnm},$ we obtain at $F=0.22\mbox{ meVnm},$
$\Omega_{0}^{\rm [tun]}\approx 0.53 \mbox{ ps}^{-1}.$ 
In the panel (a) of Fig. \ref{figsz} on can see that  at $F=0.22\mbox{ meV/nm},$
the characteristic time of spin evolution (time, at which the expectation value $\langle\sigma_{z}(t)\rangle=0$), 
is close to 45 ps, considerably larger than $\pi/\Omega_{0}^{\rm [tun]}\approx 6\mbox{ ps}.$ This difference
means that the main effect of the tunneling on the spin evolution is due to formation of the mixed states
rather than due to the driven tunneling-induced spin precession. 

By analyzing the results in Fig. \ref{figsz} - Fig. \ref{figx}, one can see that the tunneling becomes efficient at few
periods of driving field when the driving amplitude exceeds the values of about $0.18 \ldots 0.2$  meV/nm. At lower fields the 
tunneling is slow enough to allow for a well-defined spin flip. For the fields higher than $0.22$  meV/nm the potential 
well opens so effectively that the electron escapes into continuum before the well-established Rabi oscillations are developed. 
So, one needs to choose some optimal intermediate driving fields to achieve good spin flip in a proper time.
Besides, the very strong dependence of the tunneling probability on the driving field amplitude
requires a well-defined window of the field amplitude in order to achieve the desired spin flip 
if one is interested in still keeping the electron inside the hosting quantum dot.

As for the $\alpha-$dependence of the tunneling, one can see that 
the patterns of position $\langle x (t)\rangle$, the localization $P(t),$ and the energy  $\langle E (t)\rangle$ 
are similar but still quantitatively different depending on the SOC strength. For example, 
for a stronger SOC the localization probability is slightly lower at all times and for most of the driving fields, i.e. 
the tunneling, in general, becomes faster with increasing SOC. 
We attribute this effect to a more intense EDSR development at higher SOC. Stronger SOC enlarges the matrix elements for the 
states with opposite spins, leading not only to a faster spin flip but also to a larger occupation of 
higher energy levels, speeding up, to some extent, the tunneling.

\begin{figure}[tbp]
\centering
\includegraphics*[width=0.48\textwidth]{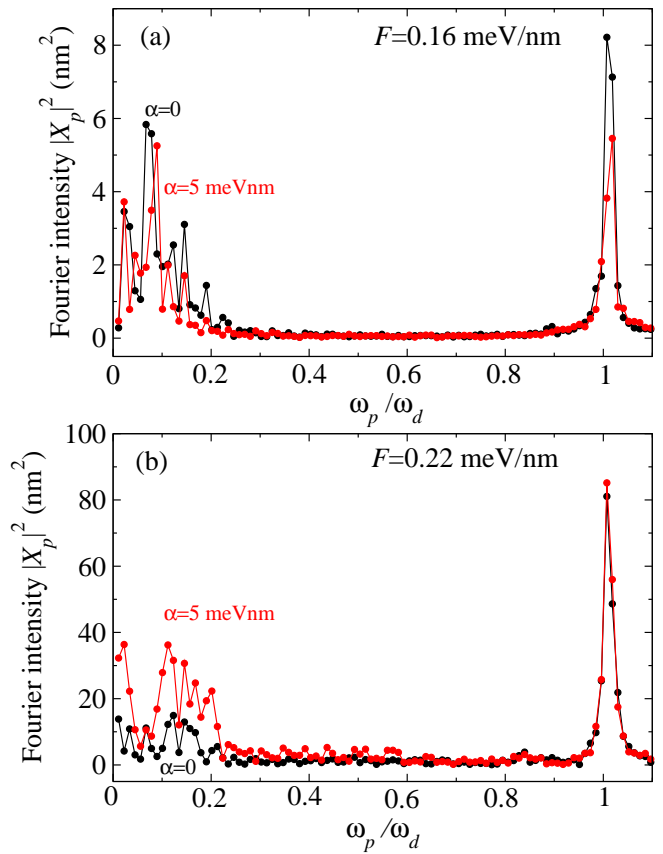}
\caption{Fourier intensities $|X_{p}|^{2}$ for $\alpha=0$ and $\alpha=5$ meVnm (as marked near the plots):
(a) $F=0.16$ meV/nm, (b) $F=0.22$ meV/nm.}
\label{figfourier2}
\end{figure}

\subsubsection{Fourier analysis of the dynamics}

It is of interest to study the driven evolution in terms of its spectral properties, namely, the Fourier 
power spectrum applied to a finite sequence of data for $\langle x(t)\rangle$ collected on the evolution 
interval $T=NT_{0}$ with $N$ points spaced by $T_{0}$. Here it is appropriate to use the discrete version of 
the Fourier transform written as:
\begin{equation}
X_{p}=\sum_{n=0}^{N-1} \langle x(nT_{0})\rangle e^{-{i}n\omega_{p}/\omega_{0}}.
\label{dpf}
\end{equation}
In (\ref{dpf}) $X_{p}$ is the output for the Fourier harmonic at frequency $\omega_{p}=\omega_{0}p/N,$ 
and (\ref{dpf}) gives Fourier data for the frequency interval 
$0 \ldots \omega_{0}/2$ with frequency spacing $\omega_{0}/N$, so the index $p$ takes the values $1 \ldots N/2,$
where it is convenient to express the frequencies in units the of $\omega_{d}$. For our numerical simulations 
we have the values $N=200 \ldots 600$, where the discrete spectra obtained from (\ref{dpf}) form a dense quasicontinous
set of states.  

Here one may expect the local maxima at both driving frequency $\omega_{d}$ and in the low-frequency part associated 
with slow spin evolution coupled to the coordinate via SOC. In Fig. \ref{figfourier2} 
we show examples of Fourier power spectra to compare the $\alpha=0$ system with the system having 
moderate SOC for driving with $F=0.16$ meV/nm and $F=0.22$ meV/nm. 
We may see that the presence of SOC modifies mainly the low frequency part of $X_{p}$
in the $\sim\,10^{-2}\mbox{ THz}$ frequency domain. This is natural due to the SOC coupling of the coordinate with spin which 
affects mainly lower frequencies associated with the spin evolution which is lower than 
the driving frequency of the electric field. The peak at $\omega_{p}/\omega_{d}=1$ is present for all SOC 
parameters and driving field amplitudes and corresponds to the dominant frequency of driven dynamics. 
Another interesting feature of Fourier spectra is the difference between the zero-
and nonzero SOC cases which increases at high driving fields as it can be seen by comparing 
panels (a) and (b) in Fig. \ref{figfourier2}. With the increase in the driving field, at $\alpha=0,$
the contribution of the large-displacement dynamics at frequency $\omega_{d}$ dominates over the low-frequency part
of the displacement spectrum. However, a finite SOC due to the anomalous velocity produces a still relatively 
intense low-frequency motion. 

It is interesting to note that in the non-relativistic nonlinear atomic optics, 
frequent recollisions of a driven electron with the singular Coulomb potential of the 
remaining ion \cite{Lewenstein} lead to a strong
enhancement in the high-frequency radiation. In our calculations we take into account 
a relativistic effect in the form of the spin-orbit couping, albeit for a nonsingular potential,
where a result of these frequent recollisions is seen as a relatively small amplitude of $\langle x(t)\rangle$
compared to $x_{0}^{\rm [cl]}$ in Eq. (\ref{xcl}), as shown in Fig. \ref{figx}(b). 
It will be of a future interest to study high-frequency harmonics produced in the presence 
of these relativistic effects.

\section{Conclusions and discussion}

We studied the electric dipole spin resonance with simultaneous tunneling and position evolution 
in a multilevel quantum dot formed in a nanowire { as caused by a driving field in the sub-THz range.} We demonstrated a strong 
effect of tunneling on the driven spin flip processes with disappearance of well-defined Rabi spin oscillations
when the tunneling rate is of the order of the corresponding Rabi frequency. For multilevel quantum dots this matching happens
in the electric field amplitudes of the order of 0.1 meV/nm, being in our realization close to 0.2 meV/nm. 
{ Thus, the tunneling induced by the driving electric field, strongly limits the maximal spin-flip Rabi frequency 
and the reliability of the spin manipulation. Although this effect is common for all systems, where electrons are localized
in finite-size potential such as donors, gate-based quantum dots, and lattice defects, every realization requires a 
system-specific analysis.} The strong effect of the tunneling on the spin dynamics is due to the formation 
of the mixed spin states, shifting the spin vector inside the Bloch sphere.   
In addition, we demonstrated 
backaction of the spin dynamics on the tunneling probability and position of the electron. This backaction 
is qualitatively attributed to the anomalous spin-dependent velocity, proportional to the spin-orbit 
coupling strength. These effects should be taken into account in the development of techniques for fast operations  
in the electron spin-based qubit architecture in semiconductor nanostructures.

\begin{figure}[tbp]
\centering
\includegraphics*[width=0.48\textwidth]{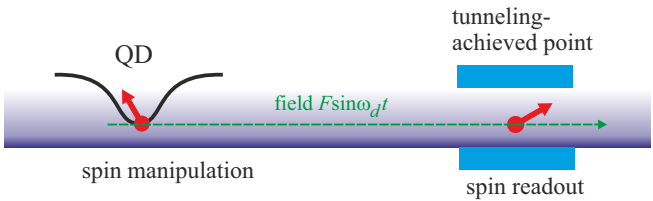}
\caption{Sketch of a device utilizing the joint effect of spin rotation and tunneling to the continuum. 
At $t=t_{0}$ the EDSR spin manipulation starts, and at some $t_{1}>t_{0}$ the readout begins for the electron outside 
of the dot. The readout process is be performed outside of the QD, thus reducing the perturbation effects of the measurement 
on the electron spin rotation in the dot and, in turn, of the spin rotation on the readout.}
\label{figdev}
\end{figure}

Finally, we would like to briefly discuss possible device implementation 
of the above predicted effects. A sketch of a device utilizing the 
tunneling to the continuum accompanying the spin rotation is shown in Fig.\ref{figdev}. 
The EDSR spin manipulation for electron in the QD starts at $t=t_{0},$ and at $t=t_{1}$ the 
readout begins at the readout area for the electron tunneled to continuum as shown in Fig.\ref{figdev}. 
At the arrival at the readout point, the electron has the spin 
components determined both by the spin manipulation in the dot and by the propagation to the readout point.
The latter can be accurately found from the knowledge of the spin precession in the nanowire including the spin precession length discussed in Sec.IV.A.
Thus, the abilities of spin manipulations can be enhanced by the tunneling process.
The readout process shown in Fig.\ref{figdev} 
is performed outside of the QD area due to the delocalization of the   
electron wavefunction, thus reducing the mutually perturbing effects of the EDSR spin manipulation and spin measurement. 
This is illustrated by Fig.\ref{mixed} where the sizable out-of-the-dot contribution 
of the wavefunction is visible and the corresponding spin components can be detected with high fidelity. 
Such setup is different from the setup using  the second QD required for spin readout in the 
EDSR experiments \cite{Nowack2007,Nadj2010} where the Pauli spin blockade has been utilized to 
enhance or suppress the current through the whole structure indicating the spin flip. 
We expect that our scheme with spatially separated spin rotation and 
readout intervals can provide alternative ways for experimental spin manipulation setups and their applications.

\section*{Acknowledgements}

D.V.K. and E.A.L. are supported by the Ministry of Science and Higher Education of the Russian Federation through the State Assignment No 0729-2020-0058. 
E.A.L. was supported by the grant of the President of the 
Russian Federation for young scientists MK-6679.2018.2.
E.Y.S. acknowledges support by the Spanish Ministry of Science, Innovation and Universities,
and the European Regional Development Fund FEDER through Grant No. PGC2018-101355-B-I00
(MCIU/AEI/FEDER, UE) and the Basque Country Government through Grant No. IT986-16. 
E.Y.S. is grateful to S. Studenikin for valuable discussion.

\end{document}